\documentclass[]{aa}
\usepackage{graphicx}
\usepackage[varg]{txfonts}
\usepackage{natbib}
\bibpunct{(}{)}{;}{a}{}{,} % to follow the A&A style
\usepackage[figuresright]{rotating}
\usepackage[a4paper,breaklinks]{hyperref}
\usepackage{balance}

\newcommand{\loggf}{\ensuremath{\log\,gf}}
\newcommand{\logg}{\ensuremath{\log g}}

\newcommand{\draftflag}{false}

\newcommand{\beq}{\begin{equation}}
\newcommand{\eeq}{\end{equation}}

\idline{10}{10}

\begin{document}

\title{The Galactic evolution of phosphorus
\thanks{Based on observations obtained with the CRIRES spectrograph at ESO-VLT Antu
8.2\,m telescope at Paranal, Programme 386.D-0130, P.I. E. Caffau}}

\author{
E.~Caffau\thanks{Gliese fellow},     \inst{1,2}
P. Bonifacio,  \inst{2}
R. Faraggiana, \inst{3}
M. Steffen     \inst{4,2}
}

\institute{ 
Zentrum f\"ur Astronomie der Universit\"at Heidelberg, Landessternwarte, 
K\"onigstuhl 12, 69117 Heidelberg, Germany
\and
GEPI, Observatoire de Paris, CNRS, Universit\'e Paris Diderot, Place
Jules Janssen, 92190
Meudon, France
\and
Universit\`a degli Studi di Trieste,
via G.B. Tiepolo 11, 34143 Trieste, Italy
\and
Leibniz-Institut f\"ur Astrophysik Potsdam, An der Sternwarte 16, 
D-14482 Potsdam, Germany
}
\authorrunning{Caffau et al.}
\titlerunning{The Galactic evolution of phosphorus}
\offprints{E.~Caffau}
\date{Received ...; Accepted ...}

\abstract%
%\context
{
As a galaxy evolves, its chemical composition changes and the
abundance ratios of different elements are powerful probes
of the underlying evolutionary processes.
Phosphorous is an element whose evolution has remained quite elusive until
now, because it is difficult to detect in cool stars. The infrared weak 
\ion{P}{i} lines of the multiplet\,1, at 1050-1082\,nm, are
the most reliable indicators of the presence of phosphorus. 
The availability of CRIRES at VLT has permitted access to this
wavelength range in stellar spectra.
}
%\aims
{
We attempt to measure the phosphorus abundance of twenty cool stars in
the Galactic disk.
}
%\method 
{
The spectra are analysed with one-dimensional model-atmospheres
computed in Local Thermodynamic Equilibrium (LTE).
The line formation computations are performed assuming LTE.
 }
%\results
{The ratio of phosphorus to iron behaves similarly to sulphur,
increasing towards lower metallicity stars. Its ratio with respect to 
sulphur is roughly constant and slightly
larger than solar, [P/S]=$0.10\pm 0.10$.
}
%\conclusions
{We succeed in taking an important step towards the understanding of the
chemical evolution of phosphorus in the Galaxy.
However, the observed
rise in the P/Fe abundance ratio is steeper 
than predicted by Galactic chemical 
evolution model model developed by Kobayashi and collaborators. 
Phosphorus appears to evolve differently
from the light odd-Z elements sodium and aluminium.
The constant value of [P/S] with metallicity implies that P production
is insensitive to the neutron excess, thus processes other than
neutron captures operate. We suggest that 
proton captures on $^{30}$Si and $\alpha$ captures on 
$^{27}$Al are possibilities to investigate.
We see no clear distinction between our results for stars with planets and stars
without any detected planet.
}
\keywords{Stars: abundances -- Stars: atmospheres -- Line: formation --
  Galaxy: evolution -- Galaxy: disk -- radiative transfer}
\maketitle

%%%%%%%%%%%%INTRODUCTION%%%%%%%%%%%%%%%%%%%%%%%%%%%%%%%%%%%

\section{Introduction}

Phosphorus is an abundant element in the Universe, and
in the solar photosphere it is among the top twenty most abundant elements. 
In the periodic table of elements, phosphorus,
with Z=15, is in the same group as nitrogen, lying between
silicon (Z=14) and sulphur (Z=16). 
Silicon is a well-studied element, and its abundance is
easily measured in photospheres  of stars of type F, G, or K, 
by analysing the \ion{Si}{i} lines.
In the last fifteen years, sulphur 
has been systematically investigated by several groups that analysed the
few multiplets of \ion{S}{i} available in the observed spectra
(for further details see \citealt{spite11} and references therein).
This is not the case for phosphorus, that, before this work, had
never been analysed systematically in cool stars.
The reason why was already given by \citet{struve30}: no \ion{P}{i} line 
is available in the ``ordinary'' range of the observed spectra of stars
of spectral type F, G, or K. Some \ion{P}{ii} and \ion{P}{iii}
lines are observable in the spectra of B-type stars.
In the UV spectrum, some lines of  \ion{P}{iv}, \ion{P}{iii},
and \ion{P}{ii} can be detected in hot stars. 
To study the variation in the phosphorous abundance over time  
we need to use long-lived stars of type F, G, or K.
In the era of the high resolution spectrograph CRIRES-VLT,
some lines of \ion{P}{i}, the lines of Mult.\,1, at 1050-1082\,nm,
can be observed. For the first time, it has thus become possible to trace the 
Galactic evolution of phosphorus.

One single stable isotope of phosphorus, $^{31}$P, 
is believed to be formed via neutron capture, as for the
parent nuclei $^{29}$Si and $^{30}$Si, probably in
the carbon and neon burning shells during the late stages of the evolution
of massive stars. The $^{31}$P produced in this way is then released
by the explosion of these massive stars as type II SNe.
\citet{woosley95} expect no significant production of phosphorus
during the explosive phases.

The handful of observations available of P in different stars
was summarised in \citet{caffau07}. Since then,
\citet{hubrig09} have derived the P abundance in horizontal branch (HB) stars in
globular clusters 
NGC 6397 and NGC 6752, where it appears to be strongly enhanced, by more
than 2 dex and almost 3 dex, respectively.
Chemical anomalies in HB stars are usually attributed to diffusion
effects, but the current models by \citet{michaud08} 
do not seem to account for more than
about 0.7\,dex of the P enhancement in the photosphere of 
HB stars of similar temperature. 

\citet{melendez09} measured P in ten
solar twins, although they do not state which lines they used.
The lines of Mult.\, 1 were not available in their observations
and we speculate that they used the strongest  line
of Mult.\,2 at 979.6\,nm.

A result concerning P evolution
that we find intriguing, is that of 
\citet{sbordone09},
who found that the sulphur abundance in the globular cluster 47 Tuc displays a 
significant correlation with the Na abundance. 
The authors point out that S 
could be produced through proton capture reactions on P
($^{31}{\rm P}\left( p,\gamma\right)^{32}{\rm S}$).
However, P in the Sun is roughly 1.7 dex less abundant than S. 
If the same abundance ratio held
elsewhere, it would be impossible to 
produce any significant amount of S at the expense of P.
The P abundance
has never been measured in stars of 47 Tuc.

In this paper, we use high-resolution
infrared (IR) spectra, obtained with CRIRES at VLT, to
analyse the behaviour of the P abundance as a function of 
metallicity in the Galactic disc.
We selected a sample of stars from
\citet{eczolfo} for which the S abundance is known, to assess 
whether the proton capture on P is a viable production channel for S. 
The sample spans a metallicity range 
of 1.2\,dex, $-1.0<{\rm\left[Fe/H\right]}<+0.3$. 
Our future plan is to extend the sample of metal-poor stars, 
and to include metallicities lower than [Fe/H]\footnote{
[X/H]$=\log\left({\rm N_X/N_H}\right)-\log\left({\rm N_X/N_H}\right)_\odot$
and [X/Y]=[X/H]-[Y/H]} =--1.

%%%%%%%%%%%%%%%%%%%%%%%%%%%%%%%%%%%%%%%%%%%%%%%%%%%%%%%%%%%%%%%%%%
\section{Observed spectra}

The strongest of the \ion{P}{i} lines (1058.1569\,nm) has an equivalent width
of about 2.2\,pm in the spectrum of the solar photosphere. 
In a solar temperature star, all the lines of Mult.\,1 are hardly
visible when the metallicity is below [Fe/H]=--0.5. 
They are of high excitation energy (see Table\,\ref{irpline}) 
and sensitive to temperature, hence become weaker in cooler stars. 
We wished to derive P abundance across as wide a range in metallicity
as possible, to study the Galactic evolution of phosphorus.
From the sample of \citet{eczolfo}, that is stars for which the abundance of
sulphur was known, we selected 22 main-sequence or turn-off, F to early 
G-type stars.
We gave preference to warmer stars, and tried to have a 
significant number of stars with detected planets.
All the stars of the sample happen to be bright, in the range 3.3-7\,mag
in J. At the VLT-Antu 8\,m telescope, these stars can be observed in twilight 
and under very poor weather conditions.

All our spectra were observed by nodding along the slit. 
The slit width was 0\farcs{2}, providing
a spectral resolution R=100\,000; the adaptive optics correction
was computed on axis, using the target star itself.  
Twenty of our stars have been observed in service mode in the program 
386.D-0130 (PI, E. Caffau).
We used the setting centred at 1059.5\,nm in order 54, 
to ensure that the four detectable lines of \ion{P}{i}
were positioned on detectors 2 and 3, which are the most efficient.
The two reddest lines of Mult.\,1 (1068.1 and 1081.3\,nm), 
that are the weakest lines of the multiplet 
in the solar spectrum, fall outside the range of the CRIRES detectors.
Our analysis was performed with the reduced spectra provided by ESO.

%%%%%%%%%%%%%%%%%%%%%%%%%%%%%%%%%%%%%%%%%%%%%%%%%%%%%%%%%%%%%%%%%%
\section{Model atmospheres and atomic data}

We analysed four of the IR \ion{P}{i} lines of Mult.\,1.
The atomic data we used, listed in Table\,\ref{irpline},
are from \citet{berzinsh97}. 

\begin{table}
\caption{Infrared phosphorus lines analysed in this work.}
\label{irpline}
\begin{center}
\begin{tabular}{rrrrr}
\hline
\noalign{\smallskip}
 $\lambda$ & Transition & E$_{\rm low}$ & \loggf  \\
 {[nm]}    &            & {[eV]}\\
\noalign{\smallskip}
\hline
\noalign{\smallskip}
 1051.1584 & 4s$^4$P$_{1/2}$--4p$^4$D$^0_{3/2}$ & 6.94 & --0.13 \\
 1052.9522 & 4s$^4$P$_{3/2}$--4p$^4$D$^0_{5/2}$ & 6.95 &   0.24 \\
 1058.1569 & 4s$^4$P$_{5/2}$--4p$^4$D$^0_{7/2}$ & 6.99 &   0.45 \\
 1059.6900 & 4s$^4$P$_{1/2}$--4p$^4$D$^0_{1/2}$ & 6.94 & --0.21 \\
\noalign{\smallskip}
\hline
\end{tabular}
\end{center}
\end{table}

For each star, we computed a 1D-LTE model atmosphere with the code ATLAS\,12
in its Linux version \citep{kurucz05,sbordone04,sbordone05},
with the parameters (${\rm T}_{\rm eff}$/\logg/[Fe/H]) from \citet{eczolfo} and 
references therein, used to derive the sulphur abundance.
The only exception is HD\,1461 for which we choose the atmospheric
parameters of \citet{sousa} and the S abundance of \citet{gonzalez10}
(see Sect.5 for details).
This choice was driven by the need 
to make a consistent comparison between phosphorus and sulphur.
The ATLAS\,12 models were computed with a mixing length parameter of 
$\alpha_{\rm MLT}=0.5$ and the overshooting option switched off.

%%%%%%%%%%%%%%%%%%%%%%%%%%%%%%%%%%%%%%%%%%%%%%%%%%%%%%%%%%%%%%%%%%
\section{Analysis}

\begin{figure}
\begin{center}
\resizebox{\hsize}{!}{\includegraphics[draft = \draftflag, clip=true]
{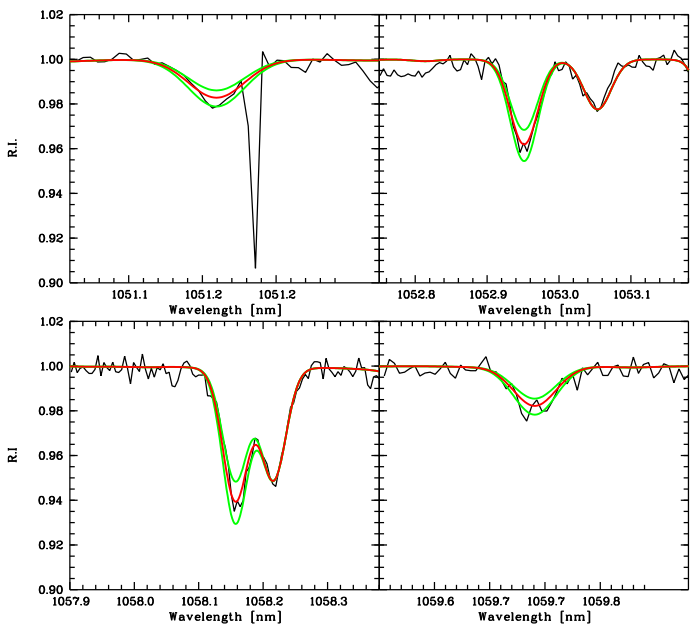}}
\end{center}
\caption[]{The four \ion{P}{i} lines are here shown
in the case of HD\,13555.
The observed spectrum (solid black) is compared to the synthetic
profile (solid red) with the P abundance derived from the EW
measurement. Also shown are the synthetic profiles obtained by
changing the P abundance by $\pm 0.1$\,dex (solid green).
The telluric absorption affecting the 1051\,pm line is clearly visible
in the upper-left panel.
}
\label{plothd13555}
\end{figure}

We measured the equivalent width (EW) of the \ion{P}{i} lines, by using the 
{\tt iraf}\footnote{Image Reduction and Analysis Facility, 
written and supported by the IRAF programming group at 
the National Optical Astronomy Observatories (NOAO) in Tucson, Arizona.
http://iraf.noao.edu/}
task 
{\tt splot}.
The lines are weak: the 
strongest line has $\log\left({\rm EW}/\lambda\right)=-5.36$,
but the majority of the lines have 
$\log\left({\rm EW}/\lambda\right) \le -5.5$,
and a Gaussian profile is a good approximation
for the line profile fitting.
The \ion{P}{i} line at 1051\,nm is, for several stars, contaminated by
telluric absorption (see upper-left panel of Fig.\,\ref{plothd13555}).
When possible, we measured the EW of the \ion{P}{i} line taking into account
the presence of the telluric contamination, otherwise we rejected the line.
Two of the lines (1052 and 1058\,nm) are blended with a \ion{Ni}{i} and \ion{Si}{i}
line, respectively. According to the strength of the lines and the stellar
${\rm V}\sin{i}$, we fitted the \ion{P}{i} line alone, or took into account
the close-by line at the same time, using the deblending option of {\tt splot}.
We used the code WIDTH \citep{kurucz93,kurucz05,castelli05} 
to derive the P abundance from the EW.
The results are listed in Table\,\ref{ab}, 
where the P abundance is given in the last but one column
together with the line-to-line scatter.
Figure\,\ref{plothd13555} shows the four \ion{P}{i} lines
for the star HD\,13555.

%%% FIGURE %%%%%%%%%%%%%%%%
%%% FIGURE %%%%%%%%%%%%%%%%

\begin{table*}
\caption{\label{ab}
Stellar parameters and abundances of our program stars.
}
\begin{center}
\begin{tabular}{lllrlllrrrrrll}
\hline\noalign{\smallskip}
  Target  & Teff & Logg & [Fe/H] & [S/H] & Ref &   S/N    & EW [pm]    &  EW [pm]   & EW [pm]   &  EW [pm]  & [P/H] & Notes\\
          &  K   &      &        &       &     & \#2/\#3 & 1051.1   & 1052.9   & 1058.1   & 1059.6   &      &\\
\hline\noalign{\smallskip}
 HD 1461   & 5765 & 4.38 & $+0.19$  &$-0.05$  & G10  &  180/150 & $    $ & $2.00$ & $2.70$ & $0.95$ & $+0.14\pm 0.01$ &P\\
 HD 13555  & 6470 & 3.90 & $-0.27$  &$-0.25$  & T02  &  400/300 & $0.80$ & $1.80$ & $2.90$ & $0.84$ & $-0.28\pm 0.03$ & \\
 HD 25704  & 5792 & 4.20 & $-0.91$  &$-0.71$  & C05  &  300/150 & $0.30$ & $0.65$ & $1.20$ & $0.30$ & $-0.55\pm 0.05$ & \\
 HD 33256  & 6440 & 3.99 & $-0.37$  &$-0.30$  & T02  &  350/250 & $    $ & $1.90$ & $3.20$ & $1.00$ & $-0.20\pm 0.04$ & \\
 HD 69897  & 6227 & 4.20 & $-0.50$  &$-0.34$  & C02  &  120/50  & $0.83$ & $2.00$ & $3.00$ & $    $ & $-0.17\pm 0.06$ & \\
 HD 74156  & 6112 & 4.34 & $+0.16$  &$-0.13$  & E04  &  250/90  & $1.15$ & $2.40$ & $3.60$ & $1.00$ & $+0.05\pm 0.04$ &P\\
 HD 75289  & 6143 & 4.42 & $+0.28$  &$-0.03$  & E04  &  180/80  & $1.50$ & $2.70$ & $3.60$ & $    $ & $+0.16\pm 0.01$ &P\\
 HD 84117  & 6167 & 4.35 & $-0.03$  &$-0.10$  & E04  &  150/70  & $    $ & $2.40$ & $4.00$ & $    $ & $+0.09\pm 0.07$ & \\
 HD 91324  & 6123 & 3.95 & $-0.60$  &$-0.49$  & C02  &  300/140 & $    $ & $1.60$ & $2.40$ & $    $ & $-0.30\pm 0.02$ & \\
 HD 94388  & 6379 & 3.96 & $+0.07$  &$+0.12$  & C02  &  200/120 & $1.40$ & $3.40$ & $4.40$ & $    $ & $+0.03\pm 0.07$ &V\\
 HD 120136 & 6339 & 4.19 & $+0.23$  &$+0.05$  & E04  &  250/170 & $1.90$ & $4.60$ & $    $ & $1.90$ & $+0.26\pm 0.08$ &P,V\\
 HD 139211 & 6231 & 4.12 & $-0.26$  &$-0.16$  & C02  &  300/240 & $1.40$ & $3.20$ & $3.50$ & $1.40$ & $+0.03\pm 0.05$ & \\
 HD 207129 & 5910 & 4.42 & $ 0.00$  &$-0.05$  & E04  &  400/250 & $0.75$ & $1.60$ & $2.10$ & $0.70$ & $-0.09\pm 0.02$ &T\\
 HD 207978 & 6400 & 4.03 & $-0.63$  &$-0.53$  & T02  &  300/170 & $0.69$ & $1.40$ & $1.60$ & $    $ & $-0.47\pm 0.05$ & \\
 HD 209458 & 6117 & 4.48 & $+0.02$  &$-0.20$  & E04  &  200/100 & $    $ & $2.00$ & $3.00$ & $    $ & $ 0.00\pm 0.03$ &P\\
 HD 213240 & 5984 & 4.25 & $+0.17$  &$-0.10$  & E04  &  300/180 & $1.40$ & $2.75$ & $3.60$ & $1.20$ & $+0.15\pm 0.01$ &P\\
 HD 215648 & 6158 & 3.96 & $-0.24$  &$-0.19$  & C02  &  220/100 & $0.96$ & $2.10$ & $2.80$ & $    $ & $-0.19\pm 0.02$ & \\
 HD 216385 & 6300 & 3.91 & $-0.27$  &$-0.18$  & T02  &  400/180 & $1.10$ & $2.40$ & $3.00$ & $0.90$ & $-0.18\pm 0.02$ & \\
 HD 216435 & 5938 & 4.12 & $+0.24$  &$+0.10$  & E04  &  250/130 & $1.80$ & $3.30$ & $4.10$ & $1.50$ & $+0.26\pm 0.02$ &P\\
 HD 222368 & 6178 & 4.08 & $-0.13$  &$-0.07$  & C02  &  280/130 & $1.30$ & $2.40$ & $3.50$ & $1.20$ & $-0.03\pm 0.02$ &V\\
\noalign{\smallskip}\hline\noalign{\smallskip}
\end{tabular}
\end{center}
The column ``Ref'' refers to the reference for the stellar parameters and
[S/H], and the corresponds to, 
G10: \citet{gonzalez10}; T02: \citet{takada02}; C05: \citet{eczolfo};
C02: \citet{Chen}; E07: \citet{ecuvillon04}.\\
The column ``Notes'' reports on the characteristics of the stars;
P: star with planets, V: variable star; T: pre-main sequence T-Tauri.
\end{table*}

We looked at the effect that an uncertainty in temperature and gravity
has on the P abundance determination.
The computations refer to the model 6112\,K/4.34/+0.16.
As a typical value for the uncertainty in the temperature 
and the gravity, we took $\Delta {\rm T_{\rm eff}}=100$\,K
and $\Delta {\rm \logg}=0.2$\,dex, respectively. Looking at the
literature, we found  that ${\rm T_{\rm eff}}$ and \logg\ in these stars
usually do not differ between different authors by more than the 
uncertainties we assume here.
The change in temperature $\Delta {\rm T_{\rm eff}}=\pm 100$
induces $\Delta\left[{\rm P/H}\right]=^{-0.03}_{+0.04}$\,dex.
For the gravity, $\Delta {\rm \logg}=\pm 0.2$\,dex 
induces $\Delta\left[{\rm P/H}\right]=^{+0.06}_{-0.05}$\,dex.
The choice of the mixing length parameter has very little influence on the 
determination of the P abundance. A change in the mixing length parameter, 
from 0.5 to 1.5, increases the P abundance of +0.0045\,dex, which is, 
for the quality of the data we analyse here, absolutely negligible.
We assumed the systematic uncertainty in the P abundance to be 0.1\,dex,
by linearly adding the contribution due to a change in effective 
temperature and in gravity. We neglect the contribution due to a change
in the mixing length parameter.
The statistical uncertainty due to the EW measurement is negligible
with respect to this systematic error.

To our knowledge, no study of the deviation from local thermal equilibrium 
for phosphorus has been published, and no P model-atom, necessary for 
this computation, seems to exist in the literature. 
\citet{caffau07} studied the effects that granulation has on the P abundance
determination in the solar photosphere. These effects are very small 
(0.02, 0.03, 0.04, and 0.02 for the four \ion{P}{i} lines we analyse here) 
compared to our error estimate, and we decided not to apply any
three-dimensional (hereafter 3D) correction.
The stars of our sample span a small range in effective temperature and gravity
and our 3D-model grid \citep{ludwig09} is very coarse.
An application of the 3D corrections interpolated from the 
small number of available models would 
result in tiny shifts in the abundances, much smaller than the associated 
error of $\pm 0.1$\,dex..

%%%%%%%%%%%%%%%%%%%%%%%%%%%%%%%%%%%%%%%%%%%%%%%%%%%%%%%%%%%%%%%%%
\section{Discussion}

\begin{figure}
\begin{center}
\resizebox{\hsize}{!}{\includegraphics[draft = \draftflag, clip=true]
{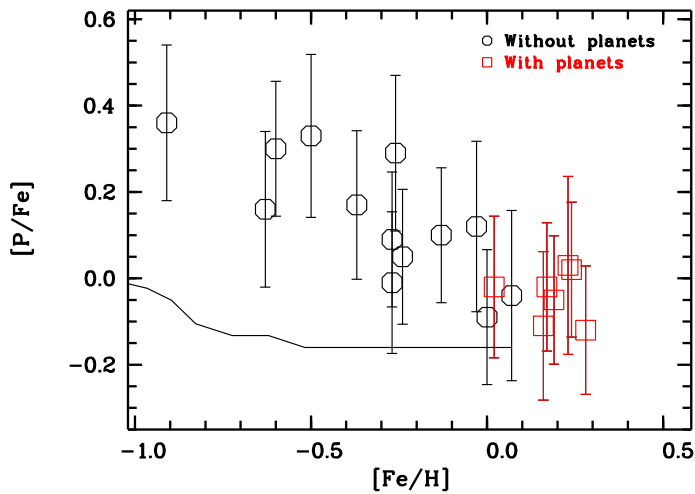}}
\end{center}
\caption[]{[P/Fe] as a function of the metallicity, [Fe/H].
The dimension of the symbols 
reflects the average line-to-line scatter of 0.045\,dex.
The error bars are the sum under quadrature
of the uncertainties of P (the linear sum of line-to-line scatter 
and the systematic uncertainty) 
and of Fe, the latter assumed to be 0.1\,dex for all stars.
Stars known to have planets (red squares) have different symbols
from stars without known planet (black hexagons).
A model of Galactic evolution of P \citep{kobayashi06} 
is added (solid line) for comparison to the observed data.
}
\label{pfevsfe}
\end{figure}

In Fig.\,\ref{pfevsfe}, [P/Fe] is plotted versus the
metallicity, [Fe/H].
The [P/Fe] ratio has a behaviour similar to that of 
an $\alpha$-element. We can clearly see [P/Fe] increases as [Fe/H] decreases
at metallicities as low as solar metallicity,
[P/Fe] is close to zero for solar metallicity stars, and increases
as metallicity decreases. 

For the light odd-Z elements it is generally assumed that
the main production channel
is neutron capture \citep{b2fh}. If this is the case,
one can expect that the production of P is sensitive
to the neutron excess\footnote{
The neutron excess is defined
by \citet{arnet71} as:
$\eta = (n_{\rm n} - n_{\rm p})/(n_{\rm n} + n_{\rm p})$, where
$n_{\rm n}$ represents the total number of neutrons per unit mass 
and $n_{\rm p}$ the  number of 
protons per unit mass.} $\eta\,$. Since the neutron excess
decreases with decreasing metallicity one can naively expect
that the ratios of P to 
elements that are insensitive to $\eta$, such as S or Fe, should
decrease with decreasing metallicity. 
The increase in [P/Fe] with decreasing metallicity contradicts
this simple expectation. 
It should however be noted that in the metallicity
range covered by our study neither Na nor Al, the nearest odd-Z
elements, display a decrease in their ratios to iron with decreasing
metallicity \citep{gehren04,shi04,gehren06}, but instead remain constant.

Owing to the small number of
metal-poor stars, it is unclear whether [P/Fe] reaches a 
plateau at about [Fe/H]$\approx -0.5$,
or continues increasing.
More data at metallicity around $-1.0$ would be very useful to clarify
this interesting question.
The observed data in Fig.\,\ref{pfevsfe} are compared to the model of
P evolution in the Galaxy by \citet{kobayashi06} (solid line).
The model seems to qualitatively reproduce the observed behaviour, 
although quantitatively the increase in [P/Fe] is stronger than predicted
and already starts at solar metallicity.

In our sample, seven stars have detected planets and are plotted 
as squares in Fig.\,\ref{pfevsfe}.
We see no difference in the behaviour in the [P/Fe] trend
between the sample with and without planets.
One should always bear in mind that
for the stars ``without planets'' it cannot be excluded that 
they harbour undetected planets. 
We note, however, that all of the stars ``without planets''
in our sample, except for HD 215648 and HD 207978
have several high precision radial velocity points measured with
the HARPS spectrograph at the ESO 3.6\,m telescope.
Given that, based on these radial velocity measurements, no planet has been
reported,  short-period massive planets (``hot Jupiters'') should
be excluded for these stars.

%%% FIGURE %%%%%%%%%%%%%%%%
%%% FIGURE %%%%%%%%%%%%%%%%

\begin{figure}
\begin{center}
\resizebox{\hsize}{!}{\includegraphics[draft = \draftflag, clip=true]
{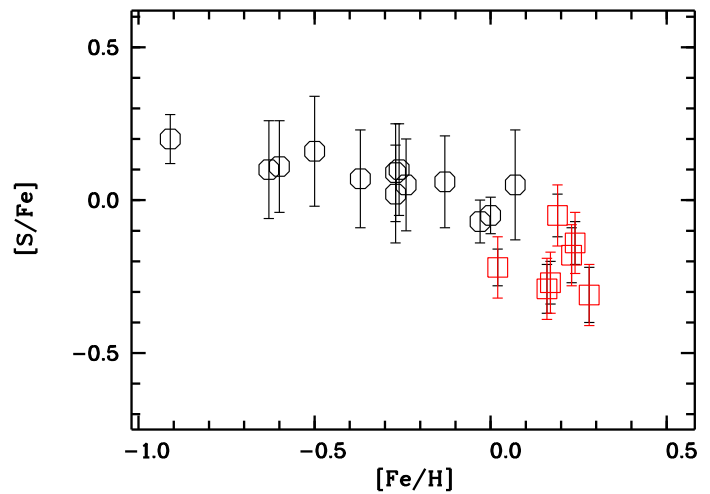}}
\end{center}
\caption[]{[S/Fe] as a function of the metallicity, [Fe/H].
The error bars indicate the uncertainties of [Fe/S] as 
given in the related papers. 
The symbols have the same meaning as in Fig.\,\ref{pfevsfe}.
}
\label{sfevsfe}
\end{figure}

\begin{figure}
\begin{center}
\resizebox{\hsize}{!}{\includegraphics[draft = \draftflag, clip=true]
{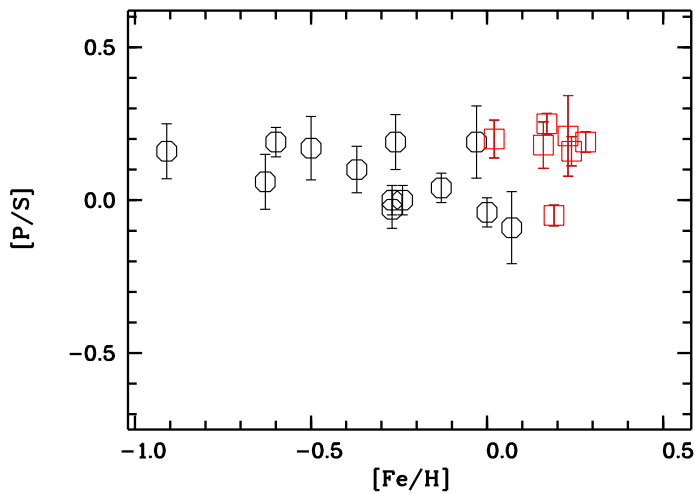}}
\end{center}
\caption[]{[P/S] as a function of the metallicity, [Fe/H].
Symbols are as in Fig.\,\ref{pfevsfe}.
The error bars are computed as the linear sum of the
systematic error of [P/S] (see text) and $\sqrt{2}$ times the
P line-to-line scatter (assuming that random errors of P and S 
are of similar size).
}
\label{psvsfe}
\end{figure}

In Fig.\,\ref{sfevsfe}, [S/Fe] versus (vs.) [Fe/H] is shown for the same sample of stars. 
There is a tiny difference between stars with and without planets. The 
difference is reflected in the [P/S] vs. [Fe/H] plot in Fig.\,\ref{psvsfe}.
Phosphorus and sulphur behave similarly when
effective temperature and gravity of the model atmosphere are changed.
We computed the systematic uncertainties in S with the models we used
for the P analysis, and assume that the systematic error in the [P/S]
ratio is given by the difference between the (logarithmic) systematic errors
in P and S.
The stars with planets are the most metal-rich ones in the sample, and have a 
lower sulphur abundances than those without planets.
The trend of [P/S] with the metallicity is flat,
and has a large scatter (see Fig.\,\ref{psvsfe}).
The average value of this ``plateau''
is [P/S]=$0.10\pm 0.10$.
As noted above when discussing the [P/Fe] ratio, this
``plateau'' in [P/S] is surprising. In the same 
metallicity range, the ratios of either Na or Al to the 
$\alpha$ element Mg decrease with
decreasing metallicity \citep{gehren06}.

For the seven stars with known planets, we find
$\langle\left[{\rm P/S}\right]\rangle=0.15\pm 0.10$, while for the stars without
known planets $\langle\left[{\rm P/S}\right]\rangle=0.07\pm 0.10$.
This finding is not statistically significant, because the two values
agree to within one $\sigma$. We also note that there is hardly any overlap
in metallicity between the stars with and without planets. Nevertheless,
this difference is puzzling and caused by the evolution in the abundance 
of sulphur, not phosphorous.

The two suspected binary stars (HD\,33256, HD\,69897) 
do not behave very differently from the complete sample. 
We can therefore conclude that the companion 
is either of the same spectral type or much fainter.
In addition, the three stars classified as variable in the
Simbad\footnote{http://simbad.u-strasbg.fr/simbad/} astronomical database 
(HD\,94388, HD\,120136, HD\,222368) and the pre main-sequence
star (HD\,207129) do not display any difference from
the other stars in the sample.
For the three variable stars, the photometric variability
of HD 94388 and HD 222368 was reported by 
\citet{petit90}.  
Later, however, these two stars were used as standards by several authors 
(e.g. HD 94388 for delta Scuti stars by 
\citet{hintz98}, 
and HD 222368 for long-term variables by 
\citet{manfroid95}. 
The slight variability of HD 120136 is studied by 
\cite{hall09},
who confirm the long-term variability detected by 
\citet{lockwood07}. 

%%% FIGURE %%%%%%%%%%%%%%%%
%%% FIGURE %%%%%%%%%%%%%%%%

%%% FIGURE %%%%%%%%%%%%%%%%
%%% FIGURE %%%%%%%%%%%%%%%%

For HD\,1461, we report in Table\,\ref{ab} 
the measurement of the sulphur abundance by \citet{gonzalez10}, who used
the atmospheric parameters of \citet{sousa}.

\citet{rivera10} found a super-Earth on a short-period
orbit (5.77 days) around HD\,1461.
This star, and its planet, has been the subject of an intense debate
between \citet{gonzalez10} and \citet{ramirez10}.
The debate revolves around the claim made by
\citet{melendez09} that the Sun has a peculiar
chemical composition with respect to the average
of solar-twins and solar analogs, being overabundant
in volatile elements and under-abundant in refractory
elements. \citet{melendez09} proceed to identify
the reason for this ``peculiarity'' is
the presence of rocky planets, such as the Earth.
This claim is supported by  \citet{ramirez10}, but rejected
by  \citet{gonzalez10} who, in an independent analysis,
do not confirm the results of  \citet{melendez09} and point
out that HD\,1461, hosting a super-Earth, has an abundance
pattern that differs from what is claimed to be a signature
of the presence of terrestrial planets by
\citet{melendez09} and \citet{ramirez10}. 
This conclusion
is contested by \citet{ramirez10}, who claim, when
restricting the analysis to refractory elements, that
HD\,1461 conforms to their claim.
In this debate, the abundance of phosphorous provides data for a
new element. In the sample of solar twins of   
\citet{melendez09}, the mean [P/Fe] is $-0.051$ with a $\sigma$ of
0.063, in excellent agreement with what we found
in HD\,1461 ($-0.050$), despite being rocky planet.

%%%%%%%%%%%%%%%%%%%%%%%%%%%%%%%%%%%%%%%%%%%%%%%%%%%%%%%%%%%%%%%%%%
\section{Conclusions.}

We have determined the phosphorus abundance in twenty stars of 
the Galactic disk that we observed with CRIRES at VLT.
The metallicity of the sample extends over about 1.2\,dex, providing
a first insight into the evolution of phosphorus in the Galaxy.
The only published model of the chemical evolution of the Galaxy that 
provides information about the evolution of phosphorus is that of \citet{kobayashi06}.
The model predicts correctly, from a qualitative point of view, the rise
of the P/Fe ratio with decreasing metallicity.
The observed rise is, however, larger by almost a factor
of two than the model's prediction, and lacks the ``plateau''' at 
metallicities [Fe/H] $> -0.5$.
Our analysis indicates that phosphorus behaves in a similar way to
an $\alpha$-element, such as sulphur.
The model of Galactic chemical evolution described by \citet{Gibson} 
predicts a $\sim 0.2$\,dex enhancement of the [P/Fe] ratio 
over two dex in metallicity, from solar to $-2.0$.
The level of enhancement at low metallicity is compatible
with what we observe, but no mention is made of the decrease
in this ratio near solar metallicity.

We have found that the [P/S] is roughly constant at a level of about 0.1\, dex.
However, the star-to-star scatter makes this value compatible with zero. 
This constant value implies that P and S are produced
in the same amounts over the metallicity range considered by 
our investigation. The picture is different from what is found
for Na or Al: both [Na/Mg] and [Al/Mg] decrease significantly
with decreasing metallicity. 
A possible interpretation is that
the P production is insensitive to the
neutron excess $\eta$. If this is the case, then
P should be produced by nuclear reactions other
than neutron captures.
Possibilities that should be explored are proton
captures on $^{30}$Si and $\alpha$ captures on 
$^{27}$Al. 
A corollary of this analysis is that, since in the Sun sulphur
is about a factor of 30 more abundant than phosphorus, this ratio
remains the same at all metallicities (in the range that we have examined).  
We therefore conclude that S production through proton captures on P 
is not a viable mechanism.
Although the possibility of a large 
enhancement of P abundance in globular clusters cannot be ruled out,
it seems unlikely. 

It is interesting to compare our results to phosphorus abundances
derived from measurements in damped Ly-$\alpha$ (DLA) systems at high redshift. 
These are gas-rich galaxies observed as absorption lines in
the direction of QSOs. 
\citet{Molaro} measured the P abundance in the DLA at $z=3.3901$
towards QSO 0000-2620. This system has a very low metallicity and measurements
of [P/H]=$-2.31$, [P/Fe]=$-0.27$, and [P/S]=$-0.40$.
The ratios of P to iron and sulphur are clearly different from 
what those seen in Galactic disk stars. This may be  due to the low
metallicity of the system, suggesting that perhaps also in our Galaxy,
the [P/Fe] ratio starts at low values, increases until reaching a maximum, 
and thereafter decreases towards solar metallicity. This behaviour
is predicted by the model of \citet{kobayashi06}, although
not shown in Fig\,.2, which only extends down to [Fe/H]=--1.0.
Another possibility is that this DLA galaxy has undergone
a different chemical evolution, since it has [S/Fe]=+0.13,
while at these metallicities [S/Fe]$\sim +0.4$ in the Galaxy 
\citep{eczolfo}.

\citet{Fenner} determined an upper limit of [P/S]$<+0.01$ in the DLA 
at $z=2.626$ towards FJ081240.6+320808.
This system is more metal-rich ([O/H]=$-0.44$), thus probably of metallicity
comparable to some of the stars in our sample. However, the [P/S] ratio is 
slightly lower than what we observe. 
Although this DLA system may resemble the 
Milky Way disk, it has probably experienced a different chemical evolution.

\balance

Our investigation is the first step towards an understanding of
the evolution of P in the Galaxy. It will be important to secure more data
at metallicity [Fe/H]=$-1.0$ and lower, to understand whether the [P/Fe] ratio
flattens out, continues to rise, or finally decreases towards low metallicity.
The good performance of the CRIRES spectrograph suggests that it will be possible to
observe considerably fainter targets with adequate exposure times.
A direct measurement of P in 47 Tuc or other globular clusters 
would allow us to probe the feasibility of S production
through proton capture in these objects.

%%%%%%%%%%%%%%%%%%%%%%%%%%%%%%%%%%%%%%%%%%%%%%%%%%%%%%%%%%%%%%%%%%%

\begin{acknowledgements}
We acknowledge support from the Programme Nationale
de Physique Stellaire (PNPS) and the Programme Nationale
de Cosmologie et Galaxies (PNCG) of the Institut Nationale de Sciences
de l'Universe of CNRS.
\end{acknowledgements}

\bibliographystyle{aa}

\end{document}